\documentstyle[11pt,aaspp4]{article}

\slugcomment{} 

\lefthead{Thompson, Mannucci, \& Beckwith}
\righthead{Infrared Emission-Line Survey}

\begin{document}

\title{A Narrowband Imaging Survey for\\
    High Redshift Galaxies\\
    in the Near Infrared\footnote{This work is
    partially based on observations collected at
    the European Southern Observatory, La Silla, Chile.}}

\author{D. Thompson}
\affil{Max-Planck-Institut f\"ur Astronomie, Heidelberg, Germany\\
Electronic mail: djt@mpia-hd.mpg.de}

\author{F. Mannucci}
\affil{C.A.I.S.M.I.--C.N.R., Florence, Italy\\
Electronic mail: filippo@arcetri.astro.it}

\and

\author{S. V. W. Beckwith}
\affil{Max-Planck-Institut f\"ur Astronomie, Heidelberg, Germany\\
Electronic mail: svwb@mpia-hd.mpg.de}

\begin{abstract}

A narrowband imaging survey of 276 square minutes of arc was carried 
out at near infrared wavelengths to search for emission line objects 
at high redshifts.  Most of the fields contained a known quasar or 
radio galaxy at a redshift that placed one of the strong, restframe 
optical emission lines (H$\alpha$, [O III], H$\beta$, or [O II]) in 
the bandpass of the narrowband filter.  The area weighted line flux 
limit over the entire survey was 
$3.4\times10^{-16}$\,erg\,cm$^{-2}$\,s$^{-1}$ ($3\sigma$), while the 
most sensitive limits reached 1.4$\times10^{-16}$\,erg\,cm$^{-2}$\,s$^{-1}$.  Integrating the volume covered by all four optical emission lines in 
each image yields a total comoving volume surveyed of $1.4\times10^5$\,Mpc$^3$.  Considering only H$\alpha$ emission in the K band ($2.05<z<2.65$), where 
the survey is most sensitive, the survey covered a comoving volume of $3.0\times10^4$\,Mpc$^3$ to a volume-weighted average star formation 
rate of 112\,M$_\odot$\,yr$^{-1}$ (for H$_0 = 50$\,km\,s$^{-1}$\,Mpc$^{-1}$, $\Omega_0 = 1$).  This is the most extensive near-infrared survey which 
is deep enough to have a reasonable chance at detecting strong 
line emission from an actively star-forming population of galaxies, when measured against simple models of galaxy formation.  One emission line 
candidate was identified in this survey, and subsequently confirmed 
spectroscopically.

\end{abstract}

\keywords{cosmology: observations --- early universe --- galaxies: 
formation --- infrared: galaxies}

\section{Introduction}

The great majority of galaxies began forming their stars early in 
the history of the universe at observable redshifts greater than 1.  
Of these, the stars in the spheroidal galaxies -- ellipticals, 
lenticulars, and the bulges of spirals -- are thought to be the 
oldest.  It is not known exactly how old the oldest stars are, 
although the upper age estimates (Coles \& Ellis 1994) are in 
conflict with the age of the universe in some cosmologies 
(Freedman et~al. 1994; Mould et~al. 1995).  It is also difficult 
to reconcile very old galaxies with models of structure formation, 
particularly the cold dark matter models (e.g. B\"orner, Mo, \& 
Jing 1995; Efstathiou 1995).  Identification of the first generations 
of stars and measurements of their ages are, therefore, of considerable 
interest to cosmologists.

There is some evidence that galaxy-sized structures existed early.  
Globular clusters have apparent ages of 14--16\,Gyr.  Galaxies at 
redshifts of order 1.5 have been discovered which appear to have 
evolved stars, thus implying a high redshift of formation (Cowie, Hu,
\& Songaila 1995, Dunlop, et~al. 1996).  Quasars are seen to redshifts 
of order 5 (Schneider, Schmidt, \& Gunn 1991) and their hosts represent at 
least a fraction of the galaxy population.  Damped Ly$\alpha$-absorption 
line systems, possibly indicative of galaxy disk formation, are also 
seen to redshifts above 4 (McMahon et~al. 1994; Lu et~al. 1996).  Many 
high redshift radio galaxies have been identified, and it has been 
suggested that they represent a stage in the formation of more normal 
galaxies (Meier 1976, Terlevich 1992, Haenelt \& Ress 1993).

There are some very exciting new results suggesting that we are 
beginning to identify large populations of high-redshift objects. 
Redshift surveys of complete samples of faint galaxies have turned 
up a number of interesting sources with $z > 1$ (Ellis, et~al. 1996,
Cowie, Songaila, \& Hu 1996).  As many as six potential clusters of 
galaxies at $z > 2$ have been 
recently been reported (Francis, et~al. 1996; Malkan, Teplitz, 
\& McLean 1996; Efstathiou, 1996).  A population of galaxies at $z > 3$ have 
been identified by Steidel et~al. (1996, see also Giavalisco et~al. 
1996), using a continuum selection technique.  And the redshift of the 
most distant ``normal'' galaxy known has been climbing rapidly, standing 
now at $z = 4.55$ with the spectroscopic confirmation of three galaxies 
associated with a quasar (Hu \& McMahon  1996).

The problem with observing the first epochs of star formation is that 
the signatures of these epochs are not known and are very poorly 
constrained.  It is usually assumed that the first galaxies emitted 
copiously in the Ly$\alpha$ line of hydrogen (Partridge \& Peebles 
1967; Tinsley 1977, 1978; Baron \& White 1987), though very few examples 
of Ly$\alpha$-bright galaxies have been found (Pritchet 1994) despite 
considerable effort (De Propris et al. 1993; Pritchet \& Hartwick 
1987, 1990; Thompson, Djorgovski \& Trauger 1995 (TDT95); Thompson \& 
Djorgovski 1995; and others).  Almost all the surveys for Ly$\alpha$ 
have been at optical wavelengths and are limited to redshifts below 
about 6.  Recent continuum-based surveys (Steidel \& Hamilton 1992, 
see also Steidel \& Hamilton 1993; Steidel, Pettini, \& Hamilton 1995) 
have been very successful in finding young galaxies near redshifts around 
3 by looking for Lyman-break objects using broad band filters. While these
galaxies appear to have fairly significant star formation rates (4--25\,M$_\odot$\,yr$^{-1}$, see Steidel et~al. 1996), they generally 
have weak or absent Ly$\alpha$ emission despite apparently low dust 
content.

This paper presents the results of a complementary approach to search 
for young galaxies, searching for redshifted emission lines of 
hydrogen and oxygen at wavelengths between 1 and 2.5\,$\mu$m.  The 
main advantage of this approach is that the lines are emitted at much 
longer rest-frame wavelengths than Ly$\alpha$ and, therefore, suffer 
considerably less from extinction by dust. Even small amounts of dust, 
of order 1 visual magnitude, are enough to quench Ly$\alpha$ emission
through resonant scattering off neutral hydrogen and even depress the
ultraviolet continuum without substantially diminishing the intensity 
of longer wavelength lines (e.g. Djorgovski \& Weir 1990; Charlot \& 
Fall 1993; Mannucci \& Beckwith 1995, hereafter MB95).  Most of the 
sources discovered by Steidel and his colleagues do have weak Ly$\alpha$, 
indicative of extinction or absorption intrinsic to the source.  An 
additional advantage of the infrared observations is that they can 
detect the Ly$\alpha$ line at redshifts exceeding 7.  A bright 
population of galaxies at these very high redshifts can only be 
studied in the infrared and are perhaps already indicated by the 
presence of quasars at redshifts around 5 (Turner 1991).

Our approach consists of taking deep images through narrow and broad 
band filters and looking for objects which are relatively brighter in 
the narrow filter, thus indicating a substantial flux in an emission 
line.  We chose the fields to contain objects with known 
redshifts which put optical emission lines (H$\alpha$, 
[OIII]\,$\lambda$5007\AA, H$\beta$, or [OII]\,$\lambda3727$\AA) in the 
passbands of the narrow filters.  If there is any tendency toward 
clustering, these objects will pinpoint regions of overdensity.

Variations of this approach have been used before by Boughn, Saluson, 
\& Uson (1986); Collins \& Joseph (1988); Thompson, Djorgovski, and 
Beckwith (1994); Parkes, Collins, \& Joseph (1994); Malkan, Teplitz, 
\& McLean (1995); Bunker, et al. (1995); and Pahre \& Djorgovski 
(1995, hereafter PD95), see also Mannucci, Beckwith, \& McCaughrean (1994).  
The main advance of the present work is the use 
of a wide-field infrared camera constructed specifically to undertake such 
surveys.  The present survey searches about two orders of magnitude more 
volume in the universe for signatures of emission lines in the infrared than 
previously accomplished.  What is presented below represents the largest 
area surveyed to date at sensitivities approaching the best in the 
literature. We assume a cosmology of H$_0$ = 50 and $\Omega_0$ = 1 
throughout this paper.

\section{Observations}

The target fields were selected primarily from the quasar lists of 
Veron and Veron (1993) and Hewitt \& Burbidge (1993).  Special 
emphasis was given to objects with either H$\alpha$ or [OIII] in the K 
band, which select the redshift ranges $2.1 < z < 2.6$ and $3.0 < z < 
3.6$, respectively. The first range corresponds to the peak comoving 
density of active nuclei, 
and may indicate an unusually vigorous period of galaxy formation.  
Objects in this redshift range have the [O III] and H$\beta$ lines in 
the H band and [OII] in the J band, facilitating the spectroscopic 
confirmation of their redshift. The second redshift range targets an 
earlier epoch for galaxy formation, and has the [OII] line in the H 
band. Preference was also given to fields at high galactic lattitude 
and which transit near zenith at Calar Alto or La Silla to minimize 
both galactic and atmospheric obscuration. Four fields have no 
known objects at redshifts sampled by the narrow band filters; these 
are referred to as {\it blank fields}.  A list of the 
observations is given in Table 1.

\placetable{table1}
\begin{table}
\dummytable{\label{table1}}
\end{table}

Images were taken in pairs through a narrowband filter and a broadband 
filter near the same wavelength.  A total of 30 image pairs were 
obtained with exposure times of 1-2 hours through the narrowband 
filter and 15-30 minutes through the broadband filter.  Each image 
consisted of many subimages of 1 minute integration time each. The 
different subimages were toward positions differing by approximately 
15\arcsec\ (dithering), allowing many unregistered images to be 
combined for background subtraction.  The background-subtracted 
subimages were then registered and combined to produce the final image 
pairs.  All data were obtained in background-limited conditions, and 
the weather was photometric or nearly so throughout the observations.

Most of the data were obtained at the Calar Alto 3.5m telescope, using 
the MAGIC cameras (Herbst et al. 1993) with wide-field optics.  In 
this configuration, the NICMOS3 256$^2$ HgCdTe array images a 
207\arcsec\ square field of view at $0\farcs 81$ per pixel.  
There is a slight loss in sensitivity due to the undersampling of the point 
spread function (PSF) at times of good seeing, but this disadvantage 
is more than offset by the large field of view available. These data 
were supplemented with images from the ESO/MPI 2.2m telescope on La 
Silla.  The IRAC2A and IRAC2B (Moorwood et al. 1992) cameras were used 
with lens system D.  These cameras also use NICMOS3 arrays.  In 
this configuration, they have a circular field of view of 
180$^{\prime\prime}$ (the corners are vignetted) with $0\farcs 7$ pixels.

\section{Data Reduction and Calibration}

The data were reduced with the IRAF package using the following 
procedure: 1) a separate sky frame was constructed for each image 
using temporally adjacent but spatially offset images, 2) each image
was then sky-subtracted and flatfielded using a normalized domeflat,
and 3) the flatfielded images were stacked into the final, deep images.
The sky frames were typically constructed using 6--12 (depending on 
atmospheric stability) images taken near the time of each individual 
exposure, scaled to have the same median sky level. The extremes of 
the distribution in each pixel were clipped and the remaining combined
either with averaging or taking a median.  As the fields were relatively 
devoid of bright objects in the first place, the resulting sky frames were 
quite clean.  The reduced frames (typically 180 images for a 90-minute 
series through the narrowband filters) were registered on objects visible 
in the individual images and offset using integer pixel shifts.  The 
stacking was done in a similar manner to the construction of the sky frames, 
clipping pixels at each end of the distribution (ignoring flagged bad 
pixels), and averaging the remaining values.  The final stacked images were 
then trimmed to include only the highest sensitivity regions where both 
the broadband and narrowband stacks overlap. 

Positions were determined for all of the objects on the narrowband 
images using FOCAS (Jarvis \& Tyson 1981, Valdes 1989), with the 
number of connected pixels set to 4-6, depending on the seeing, and 
the threshold for detection to 3$\sigma$.  For our data, these settings 
recover all objects within the frame that are readily apparent without
producing excessive numbers of obviously false detections. Instrumental magnitudes were determined for each object in the broadband and 
narrowband images through circular apertures using the PHOT routine 
in IRAF.  The photometric aperture was set to twice the seeing full-width 
at half-max (FWHM), which maximizes the signal-to-noise ratio for faint,
point-like sources (TDT95, Howell 1989).  Small aperture corrections were 
then applied as determined from brighter point sources in each individual 
image. This procedure will underestimate the flux from bright, extended 
sources, although it should have minimal effect on high-redshift galaxies 
which subtend only a few seconds of arc. Note that faint sources more extended 
than $\sim 2^{\prime\prime}$, corresponding to $\sim$15 kpc at $z = 2.5$,
have a lower probability of being identified by FOCAS in the first place.

The magnitude scale was established with the Elias standards (Elias et 
al. 1982) and the UKIRT faint standard star list (Casali 1992), with the 
flux density zero points taken from Bessel \& Brett (1988).  For broadband 
observations in the K$^\prime$ filter, we assumed that the standard 
stars had the same magnitude in K$^\prime$ as in the standard K 
filter.  Calibration of the narrowband images was complicated by the 
presence of 12 hydrogen white dwarf stars among the 34 UKIRT 
standards\footnote{It was also found that FS24 (SA 106-1024) is listed 
as a variable star.}.  Initially, narrowband magnitudes for the 
standards were interpolated from their broadband magnitudes, using the 
filters' central wavelengths and widths.  This method is inaccurate 
when used with hydrogen white dwarf stars, because they have strong 
absorption lines in the hydrogen line filters (e.g., Pa$\beta$ and 
Br$\gamma$).  To provide accurate calibrations, we assumed that the 
majority of objects in each frame had no excess emission line flux.  
Testing this method on photometric data with both white dwarf and 
subdwarf standards shows the method to be reliable, with an accuracy 
better than 0.1 mag.

\section{Results}

Plots of the (broad$-$narrow) color vs. narrowband magnitude were 
constructed for each of the 30 image pairs.  Objects with relatively 
strong emission lines stand well away from the locus of the 
remaining objects.  All of the known AGN showed strong emission line 
signatures when H$\alpha$ was in the narrowband filter, while the radio 
galaxy B3 0806$+$426 showed strong emission in the [O III] 
$\lambda$5007\,\AA\ line.  There was only a single strong candidate 
identified in the survey data, discussed below.  A summary of the reduced 
data (seeing, trimmed field size, and flux limits) are given in
Table 2.  Total errors on the determination of the emission-line
flux limits are estimated to be $\pm$20\%, based on multiple measurements 
of the standard stars.  

\placetable{table2}
\begin{table}
\dummytable{\label{table2}}
\end{table}

A color-magnitude diagram for the B2 0149$+$33A field is presented in Figure 
1, which illustrates the selection criteria for candidate emission-line 
objects.  The curved line indicates the 3$\sigma$ emission-line flux limit,
taking into account the uncertainties from both the broad and narrowband 
images.  In this field, the line does not go through the zero color line 
because the central wavelength of the narrowband filter is near the red end 
of the broadband filter transmission.  The curve was derived in a similar 
manner to that in Bunker et~al. (1995). The horizontal dashed line indicates 
a 50\,\AA\ restframe equivalent width limit for emission lines. This limit 
is lower than the equivalent Bunker et~al. (1995) limits mainly due to the
narrower filters used in this survey (1\% vs. 1.75\%).  The diagonal, 
long-dash line is a color limit, where objects falling to the left of this 
line are well detected in both the broad and narrowband images.  Candidate
objects must lie above the equivalent width limit and to the left 
of the $3\sigma$ flux limit.  We note that an object with an infinite 
equivalent width emission line would have a color of 3.1 (for this field,
off the top of the plot), independent of its magnitude.  Many of the 
objects with this color lay near the edges of the narrowband images in 
regions where the noise was a little higher than in the center of the 
image.  They were inspected carefully by eye, and dropped from further consideration if they appeared to be spurious detections.  Finally, the 
image pairs were ``blinked'' on a workstation monitor to determine the 
quality of the candidates by inspection.

\placefigure{fig1}	

\subsection{The B2 0149$+$33 field}

The quasar ($z = 2.431$) shows strong emission with a line flux of 
8.4$\times10^{-15}$ erg\,cm$^{-2}$\,s$^{-1}$ (labelled ``QSO'' in 
Figure 1).  The restframe equivalent width of the 
emission line, as derived from the broad and narrowband images, is 
130\,\AA.  There is a companion galaxy $2\farcs 2$ to the southwest 
at PA$=249^\circ$ which is evident in both the broad and narrowband images.  
We subtracted a point-spread-function from the quasar image to 
check for excess flux in the companion galaxy; none was found -- the 
companion is plotted with an open circle on Figure 1.  
The quasar shows a damped Ly$\alpha$ system at $z = 2.133$ (Wolfe 
et~al. 1986).  If this galaxy is the absorber, it lies 18 kpc from the 
line of sight (H$_0$ = 50, $\Omega_0$ = 1). It has a K$^\prime$ 
magnitude of 17.85.  There are several other fainter galaxies which 
lie within 15\arcsec\ of the quasar.

Object 39 in Figure 1 is the strongest emission-line candidate in this 
survey with a derived emission line flux of 6.5$\times10^{-16}$
erg\,cm$^{-2}$\,s$^{-1}$, a 10$\sigma$ emission-line detection in this 
field.  It is an extended object with a K$^\prime$ magnitude of 18.52 
lying 50$\farcs 1$ from the quasar at PA 295$\fdg 1$.  The restframe 
equivalent width for the emission line is 125\,\AA, which is high, though
not unreasonable, for nearby starburst galaxies.  If this galaxy is at 
the same redshift as the quasar, the line flux would correspond to a 
restframe line luminosity of 2.8$\times10^{43}$\,erg\,s$^{-1}$.  Using 
the empirical conversion of H$\alpha$ luminosity to star formation rate 
of Kennicutt (1983, see also MB95), the star formation rate is
250\,M$_\odot$\,yr$^{-1}$.  Preliminary reduction of spectra of this 
object, obtained in the K band at UKIRT, confirm the emission line, and 
will be discussed more extensively in a separate paper (Beckwith et~al. 
1997). The images of this field are shown in Figure 2.

\placefigure{fig2}	

Object 27 corresponds to a short, thin arc in the narrowband image,
whose width is less than the seeing FWHM. Although there is a faint
galaxy at this position in the broadband image, this candidate is
considered to be a spurious detection.

\subsection{Survey limits}

Several near-infrared searches for emission-line objects have been 
published during the last few years (TDB94, PD95, Bunker et al. 1995, 
Parkes et al. 1994) and a few others have been undertaken (Malkan, Teplitz, 
\& McLean 1995).  All make use of narrowband imaging to identify candidate 
emission line objects.  Figure 3 is a comparison of several surveys in 
the observed quantities of area vs. line flux.  The relative advantages 
of the approach adopted here is evident from this figure.  The area 
covered in the present survey is approximately two orders of magnitude 
greater than the other most recent works, those of Bunker et~al. (1995) 
and PD95. The flux limits of the present work are about four times higher
than that in PD95, but as discussed below the area coverage has significant
advantages to search for the emission signatures expected from young 
galaxies at high redshift.

\placefigure{fig3}	

\subsection{Restframe Limits}

While Figure 3 compares surveys in observed coordinates, 
the survey limits can be better expressed in restframe coordinates within 
an assumed cosmology.  Although the full parameter space is 3-dimensional 
(MB95, TDT95), with axes of comoving volume density, line luminosity, and redshift, it is generally viewed in projection along the redshift axis.  
Working in the restframe allows a direct comparison between the survey results and models of galaxy and star formation rates versus the comoving volume 
density of objects.  To calculate the total volume sampled by the survey, 
we include the five strong lines: H$\alpha$, H$\beta$, [OIII]$\lambda5007$,
[OII]$\lambda3727$, and Ly$\alpha$.  Each of these appears in the narrowband
filters at different redshifts.  Each imaged field thus covers five sampled
volumes at different redshifts.  The total volume covered by the survey in 
the four optical lines (Ly$\alpha$ is considered separately, below) is
1.4$\times10^5$\,Mpc$^3$ for $H_0 = 50$\,km\,s$^{-1}$\,Mpc$^{-1}$, 
$\Omega_0 = 1$.

Assuming a Poisson distribution (no clustering), the probability, $P_0(V)$, 
that {\it no} objects are present in the sampled volume, $V$, is 
$P_0(V) = \exp(-\rho V)$, where $\rho$ is the average density of 
objects.  Prior surveys similar to this one have calculated their volume
density simply as the inverse of the volume surveyed, giving only
a 63\% chance of detecting anything.  For this survey, we adopt a more
conservative 90\% probability of detecting {\em at least} one object in 
the sampled volume, which requires that $0.9 \leq 1 - \exp(-\rho V)$, 
implying that $\rho \geq 2.3/V$.  Both the observed frame limits plot 
(Figure 3) and the restframe limits discussed below incorporate this 
formula to correct the survey to 90\% detection probability.  All of the 
data taken from the literature and used here has been corrected for this 
factor of 2.3 as well.  

Line luminosities are converted into equivalent star formation rates by
assuming a linear correlation between the two quantities and adopting the
normalization for H$\alpha$ given by Kennicutt (1983) and assuming line 
ratios relative to H$\alpha$ as in MB95.  The survey limits on the number
density of objects with various line fluxes are then transformed into 
volume density of objects with different star formation rates at discrete
redshift intervals.  These relationships may be compared directly with
models of galaxy formation in the early universe, discussed in the next 
section.

It is important to emphasize, however, that the survey limits plotted 
in following the 4 figures assume no obscuration by dust, and therefore
represent lower limits to the true star formation rate.  Even without 
considering resonant scattering of the Ly$\alpha$ photons, simple dust
absorption, assuming a galactic extinction law and $E_{(B-V)} = 0.3$
(corresponding to $\sim 1^m$ of visual extinction), would imply that the
true star formation rates are higher than that of no obscuration by a 
factor of 2 at H$\alpha$, 2.8 at H$\beta$, 3.6 at [O II], and 16 at 
Ly$\alpha$ (Seaton 1979).  We have included in the following plots an
arrow showing the magnitude and direction that an extinction of $E_{(B-V)} 
= 0.3$ would have on the plotted survey limits.

\subsubsection{Galaxy Formation Models}

Figures 4--7 compare the survey limits 
to the expected comoving density and luminosity of forming galaxies, 
taking into account various star formation histories and mass evolution.  
The model details are described more thoroughly in MB95 and are only 
summarized here.  These models estimate the {\em minimum}
density of young galaxies as a function of star formation rate which are
necessary to produce the local population of elliptical galaxies.  The 
estimated densities depend on the normalization, $\phi^*$, of the local 
galaxy luminosity function.  We assume $\phi^*$ as given in Baron \&
White (1987), though this has subsequently been revised upwards by 
a factor of about 2.2 by the most recent redshift surveys, as in Ellis 
et~al. (1996).  Use of this higher normalization would increase our 
model galaxy densities by this factor.

We consider three classes of models, ranging from strong bursts to distributed 
star formation.  The first and simplest assumes a constant star formation 
rate for each galaxy from a starting redshift $z_{in}$ to a final redshift
$z_{fin}$.  This constant level of star formation is determined separately for each galaxy, based on its mass and the available time.  These models are referred to as {\em constant} models, and 
labelled as such in Figures 4--7.  The
boundaries correspond to star formation histories starting at an initial
redshift, $z_{in}$, and finishing at a final redshift, $z_{fin}$, chosen
to maximize or minimize the available time (i.e. 
bracket the range of reasonable star formation rates) and correspond to 
the minimum and maximum expected brightness, respectively.  
The constant models assume no evolution in comoving volume 
density and a relatively quiescent star formation history.  Galaxies which
fit these models can also result from a bottom--up formation scenario, if 
we assume the galaxies were originally divided into $N$ subsystems, each one forming stars in $1\over N$ of the available time.

In the second class, each galaxy creates all its stars in a single 
burst over a fraction of the available time from 
$z_{in}$ to $z_{fin}$.  The observed luminosity function is thus 
shifted towards brighter galaxies which have lower density in their 
bright phase relative to galaxies which use the entire interval for 
star formation.  We adopt a fractional interval of 0.2 and the same 
values of $z_{in}$ and $z_{fin}$ as in the constant models.  These 
are labelled the {\em burst} models in Figures 4--7.

The third class of models assumes that most of the star formation 
takes place in small fragments or {\em subsystems} which later merge 
to form the final galaxy.  We assume that star formation takes place 
at a constant rate over all of the available time and adopt five 
subsystems for this class.  The assumption of a constant SFR in 
relatively small fragments results in a particularly low SFR in the
subgalactic fragments, and
thus these models are pessimistic for observers.  Subdividing the galaxies 
into  fragments shifts the resulting population toward more abundant but 
less luminous galaxies relative to the constant models.  These are labelled 
the {\em hierarchical} models in Figures 4--7.
The apparent brightness of $L^*$ galaxies is indicated as a horizontal line 
on each model curve in all three figures.

It is likely that all three processes play some role in galaxy 
formation: merging is seen in the ultraluminous starburst galaxies, 
disks appear to grow continuously over Hubble timescales, and the 
high-redshift quasars imply rapid star formation at early times, 
perhaps in fragments much smaller than galaxies.  There are few, if 
any, constraints on the free parameters in such a mixed model, but the 
general trend of particular mixes can be assessed by inspection of the 
plots, weighting the different classes for any hypothesis.

To follow the cosmic evolution of model expectations and survey 
results, there is a separate figure of comoving volume density vs. 
star formation rate for four redshift ranges.  For each redshift 
range, the survey volume is restricted to that sampled by appropriate
combinations of the five bright lines and the narrow filters' bandpasses, 
so the survey limits change somewhat among the four figures.  

\subsubsection{The redshift range $2 < z < 3.5$}

This redshift range contains most of the known quasars in the fields 
and corresponds to the epoch of the maximum QSO density. At these 
redshifts, objects with unobscured star formation rates equal to 
100\,M$_\odot$\,yr$^{-1}$ would be readily detected.  The age 
of the universe was about 2.5\,Gyr at $z = 2$ in the adopted cosmology.
The maximum time available is long, and, therefore, the minimum luminosity 
is low, corresponding to a very long, quiescent phase of star formation.
There is, however, no clear upper limit to the maximum luminosity, because 
the formation timescale could be as short as desired, though this just represents a blurring of the distinction between the constant and burst 
models.

\placefigure{fig4}	

Figure 4 shows that the data sample enough volume to 
exclude the {\em constant} and {\em burst} models, although the limits 
only partially overlap the expectations for {\em hierarchical} 
models.  Because the 90\% detection probability limit line for this 
survey intersects the burst and constant models for galaxies fainter 
than $L^*$, statistical fluctuations in the (small) number of more 
massive galaxies present in any field will not change these 
conclusions.  Young galaxies can be present in this redshift range 
only if they are small systems, have a low surface brightness, or are 
less efficient in emitting lines.  Alternatively, the majority of 
galaxy formation could have occurred at higher redshifts.

The galaxies discovered by Steidel et~al. (1996) have a lower density 
and/or a lower star formation rate (indicated by the heavy dashed line 
labelled S$+$96 in Figure 4) than any of the models.  The velocity 
dispersions are similar to those seen in local spheroids and their 
density lower than the comoving density of L$^*$ galaxies found by 
Ellis et~al. (1996) indicate that these are relatively massive systems.   
With their low apparent star formation rates (with typical values of about 
8.5\,M$_{\odot}$\,yr$^{-1}$), these galaxies would not be able to form all 
of the stars in an L$^*$ galaxy in the time available ($\sim$1\,Gyr) at 
a redshift of 3.  Allowing the star formation to continue down to $z = 1.5$
and assuming a low-density cosmology would still be insufficient.  
Therefore, either a large fraction of the gas in these systems has yet to 
be converted into stars (if a constant SFR is assumed), the galaxy has
already formed most of its stars (implying much higher star formation 
rates at earlier epochs), or sufficient dust exists in these objects
to mask the true, higher star formation rates. 

\subsubsection{The redshift range $3.5 < z < 5.5$}

Figure 5 shows the redshift range corresponding to 
the appearence of the QSO population.  At these redshifts, galaxies 
with unobscured star formation rates of 200\,M$_\odot$\,yr$^{-1}$ or 
more would have appeared in the survey, which is relatively high in 
comparison to known starburst galaxies at lower redshifts.  The age of 
the universe at these redshifts is about 1\,Gyr.  The models shown here 
use $z_{fin} = 3.5$, since later star formation should be detected
preferentially by observations at the lower redshifts discussed in the 
previous section.  The onset of star formation in these models begins at
redshifts between 5.5 and 20.

At these higher redshifts, the data sample a sufficient volume to 
challenge the models: at least one galaxy should have been detected, 
if the {\em burst} models apply.  There is also a significant overlap 
with the {\em constant} models, only those with the more quiescent star 
formation histories would escape detection.  The {\em hierarchical} 
models are still fainter than the limits. As in the previous figure, the 
limit line intersects the {\em burst} and {\em constant} models above the 
L$^*$ value.

\placefigure{fig5}	

\subsubsection{The redshift range $7 < z < 20$}

Ly$\alpha$ ($\lambda$1215.7\,\AA) appears in the $JHK^\prime$ filters at redshifts between 8.3 and 18.7.  There are several regions of good atmospheric 
transmission near 1 micron, of which both this survey and Parkes et 
al. (1994) make use, extending the observable redshift range down to 
$z \sim 7.2$.  Figure 6 plots the restframe 
limits for this survey with Ly$\alpha$ in the $J$ band, $7 < z < 10$.  
These redshifts may sample the parent population of 
galaxies which become quasars by $z > 4$.  The universe at these 
redshifts was only $\sim 0.5$\,Gyr old.  Even short bursts of star 
formation, lasting of order $10^8$ yr, would span a $\Delta z > 1$, 
making surveys at these very high redshifts less sensitive to the 
exact timing of the onset of galaxy formation.  Our deepest fields can 
detect an unobscured star formation rate of 220 M$_\odot$\,yr$^{-1}$.

\placefigure{fig6}	

The $K$-band part of the survey includes a significant volume 
containing Ly$\alpha$ in the range $15.5 < z < 18.7$ 
(Figure 7).  Our most sensitive data could detect star 
formation rates of 500 M$_\odot$ yr$^{-1}$, although the data do not 
go deep enough to significantly challenge the model predictions.  They 
are meaningful if much of the star formation took place at these very high 
redshifts, in which case the star formation rates would necessarily 
be quite large.

\placefigure{fig7}	

\section{Conclusions}

Young galaxies with emission lines at infrared wavelengths are rare at 
the level of sensitivity and area coverage that can be reached with 
the current generation of detectors. This survey of 276 square arcminutes 
at a limiting sensitivity of approximately $3\times 
10^{-16}$\,ergs\,s$^{-1}$\,cm$^{-2}$ in the J, H, and K$^\prime$ bands 
revealed one emission-line object likely to be at $z = 2.43$. 

The failure to detect many young galaxies makes it unlikely that most 
galaxies had star formation histories with continuous formation 
starting at any redshift and continuing to about $z \sim 2$.  Hierarchical 
formation, in which galaxies were assembled from many pieces over a 
long interval are consistent with the results. 

These conclusions stem from a comparison of the limits of the 
survey to calculations of expected line fluxes from various galaxy 
formation models.  It is possible that physical conditions not 
included in the models could reduce the observable line strengths, and
would weaken the conclusions derived from upper limits.  Dust along the 
lines of sight to the galaxies could hide galaxies whose local line fluxes 
are above the detection limit (the effects of dust absorption are indicated 
in Figures 4--7).  Young galaxies might be very extended, 
making the surface brightness too low to see in this survey, even 
though the overall line luminosities might be large.

Nevertheless, the availability of large-format infrared detectors makes 
it possible to conduct meaningful searches for high-redshift 
galaxies.  The sensitivity reached by the present survey over the 
survey area is good enough to detect very young galaxies with a 
variety of possible star formation histories that can be considered 
``conventional.''  Increases in array sizes coupled with wide-field 
optics make it attractive to continue the search for the first 
generation of stars at infrared wavelengths. 

The combination of the sensitivity and area coverage of the present 
survey when compared with these conventional models underscores the 
importance of wide-field cameras to sample large volumes of the 
universe.  Steidel et al. (1996) find 0.4 objects per square arcminute 
between $3.0 < z < 3.5$, or $\sim 4 \times 10^{-4}$ galaxies Mpc$^{-3}$.
Several tens of such objects should be in the volume searched in the 
present work, although they would lie below the current detection limit.  
If the mass function for forming galaxies is reasonably approximated by
the local one, as assumed in MB95 and the models used here, then the 
apparent number density of galaxies does not rise steeply with decreasing 
line flux.  If these models are approximately correct, the higher 
sensitivity in the greatly restricted fields available on current or 
future large telescopes will be less effective in identifying a population 
of such objects than wide field searches.

This project is part of a continuing program at the Max-Planck-Institut 
f\"ur Astronomie to identify the formation of the first generation of 
galaxies, combining optical and infrared instrumentation to search a 
large volume of the parameter space available from the ground.

\acknowledgments

We are grateful to the team that constructed the MAGIC cameras, 
particularly T. Herbst and M. McCaughrean, for making this survey 
possible.  We thank J. Cohen, G. Efstathiou, K. Meisenheimer, A. 
Putney, M. Schmidt, C. Steidel, and S. White, for discussions which 
contributed substantially to our understanding of high redshift 
galaxies.  This research was supported by the Max-Planck-Society.

\clearpage

\begin{figure}
\plotone{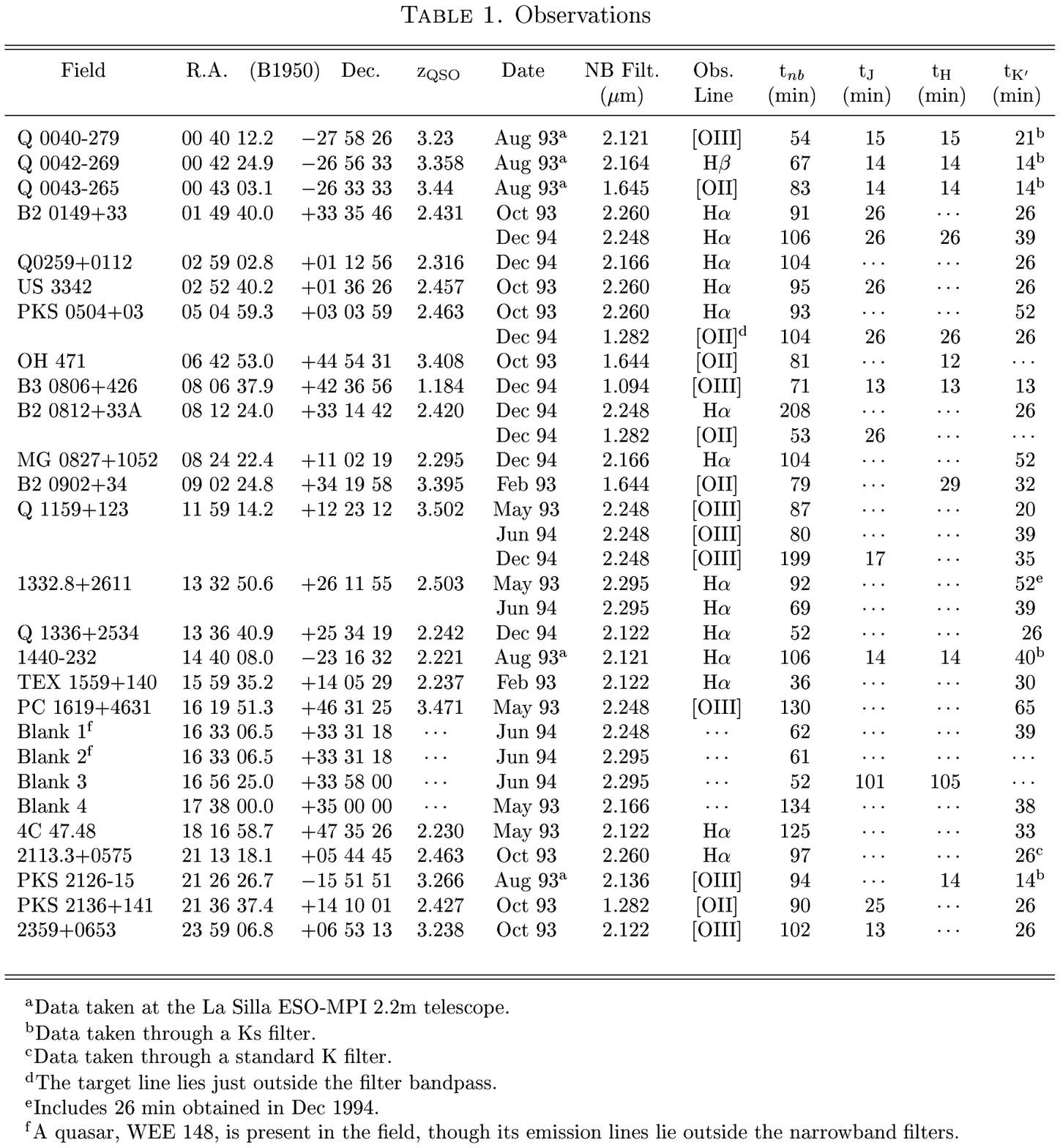}
\end{figure}
\clearpage

\begin{figure}
\plotone{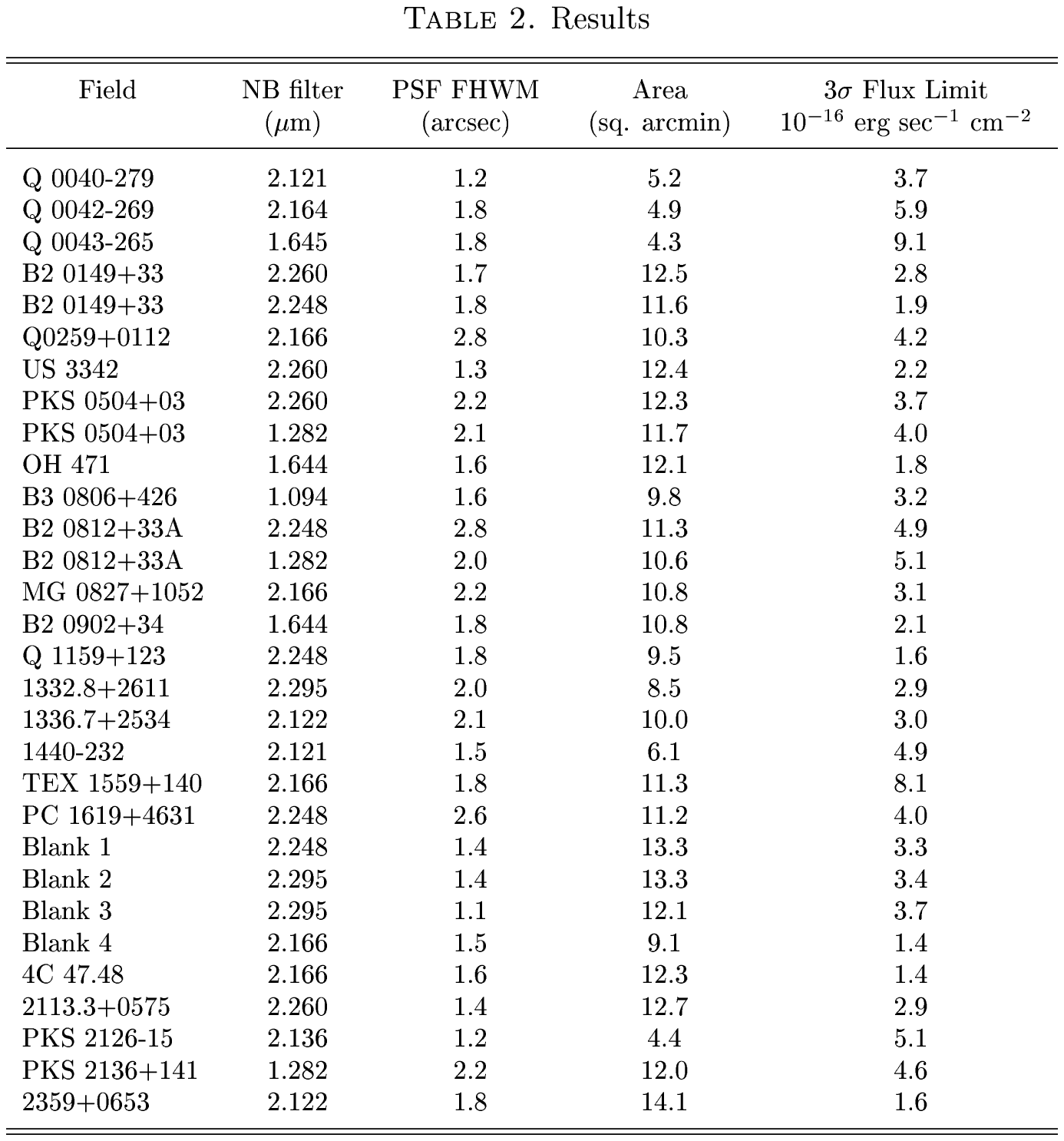}
\end{figure}
\clearpage

\begin{figure}
\plotone{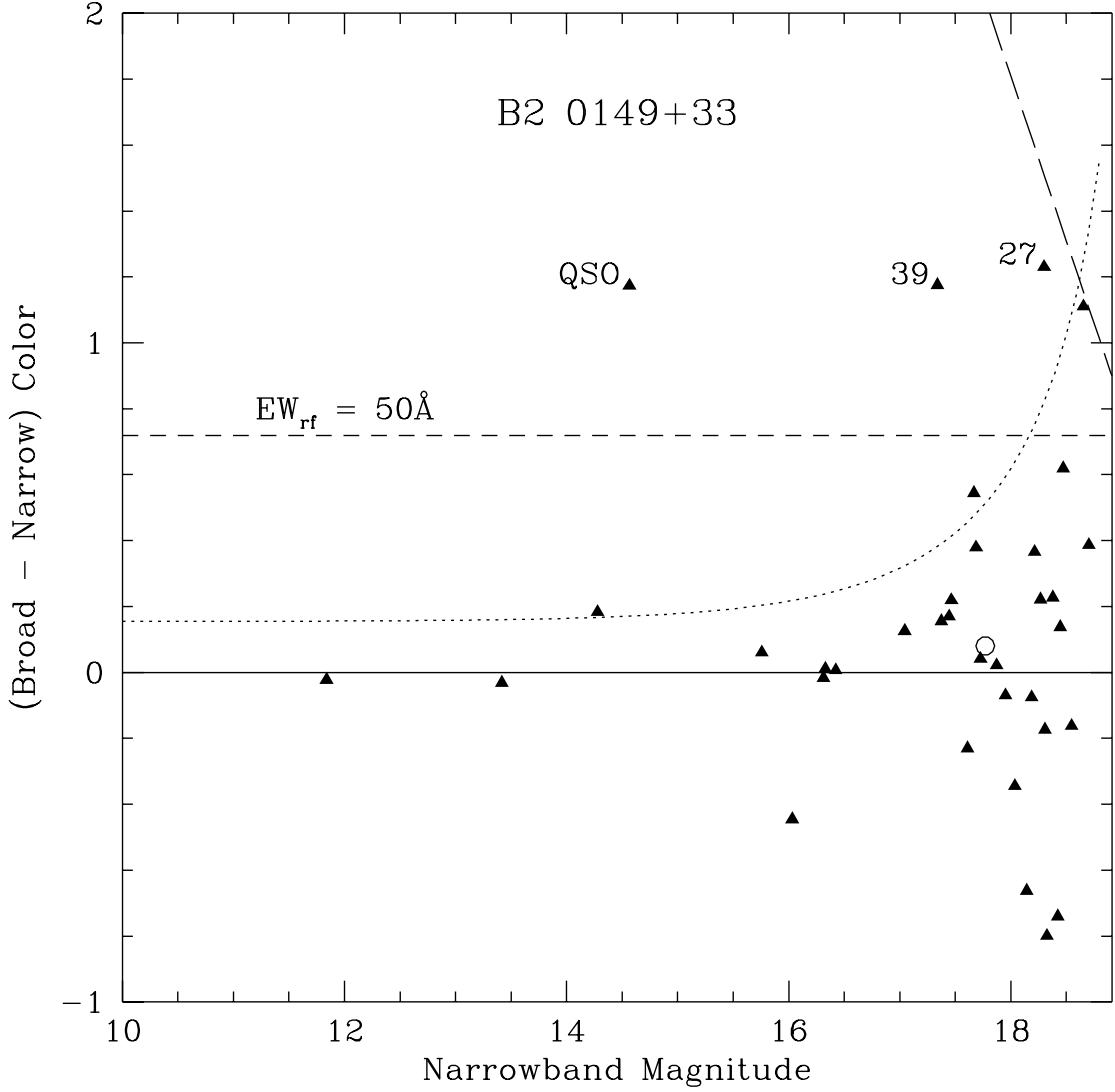}
\figcaption[Thompson.fig1.ps]{Color-magnitude diagram for the B2 
0149$+$33 field. The curved line indicates 3$\sigma$ emission line 
flux, while the horizontal dashed line is set to an emission line 
equivalent width of 50\AA\ in the restframe of the quasar. Objects 
to the left of the diagonal long-dashed line are well-detected in 
both images. The quasar and objects \#39 (the emission line candidate) 
and \#27 (see text) are labelled, while the galaxy next to the quasar 
is marked with an open circle. \label{fig1}}
\end{figure}
\clearpage

\begin{figure}
\plotone{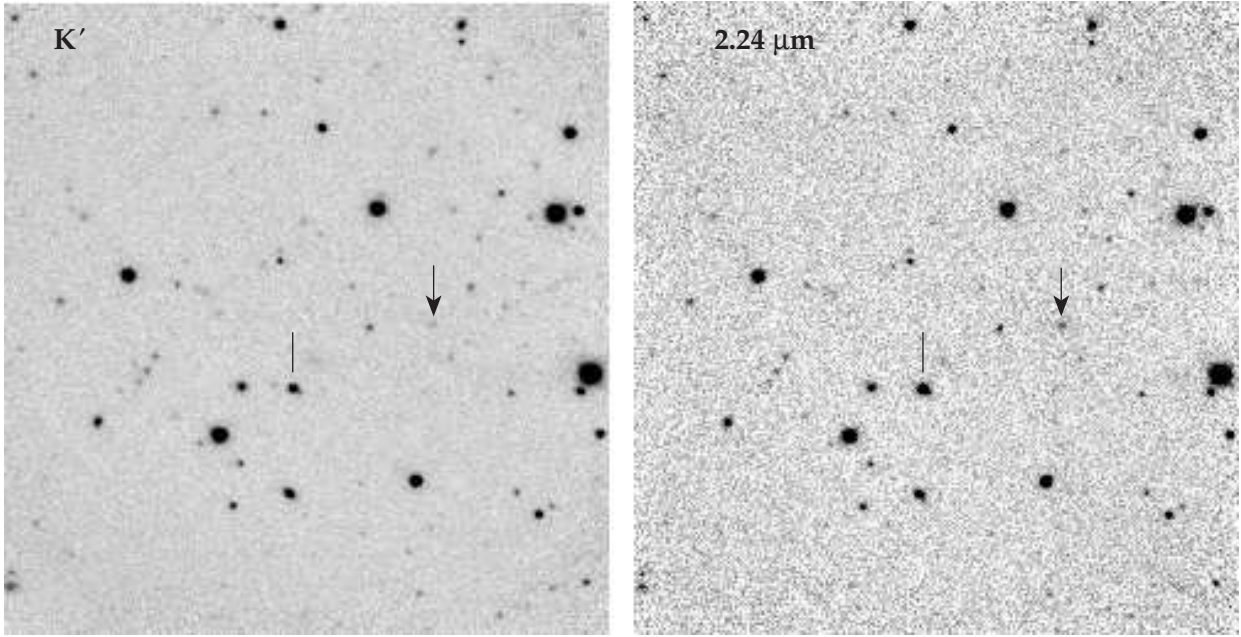}
\figcaption[Thompson.fig2.ps]{Images of the B2 0149$+$33 field through 
the broad (left) and narrowband (right) filters. The quasar (tick mark) 
and the emission line object \#39 (arrowed) are indicated. The images 
are oriented with north up and east to the left, and are scaled so that
continuum objects appear the same in both images. Emission-line objects, 
such as the quasar and object \#39, appear brighter in the narrowband 
image. \label{fig2}}
\end{figure}
\clearpage

\begin{figure}
\plotone{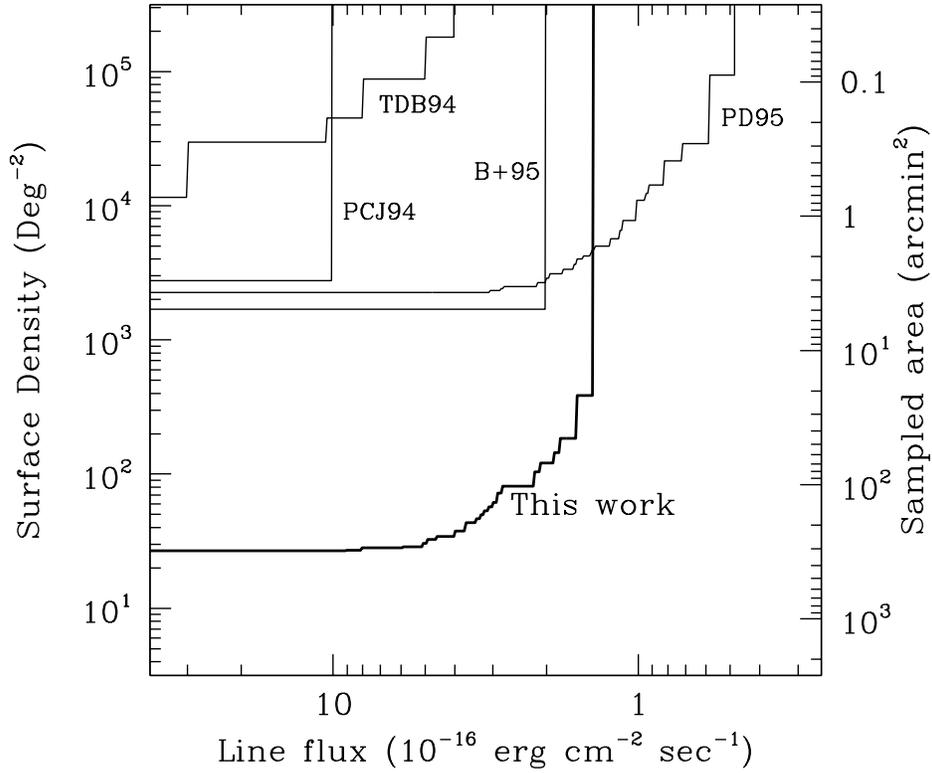}
\figcaption[Thompson.fig3.ps]{Comparison between the various published 
NIR emission-line surveys.  The vertical scale on the right is the 
sampled area, the one on the left is the minumum detectable surface 
density at 90\% confidence.  The thick line is the limit from this 
work, while the thin lines are the limits from other published work 
(PD95: Pahre and Djorgovski, 1995; TDB94: Thompson, Djorgovski and 
Beckwith, 1994; B+95: Bunker et al, 1995; PCJ94: Parkes, Collins and 
Joseph, 1994). \label{fig3}}
\end{figure}
\clearpage

\begin{figure}
\plotone{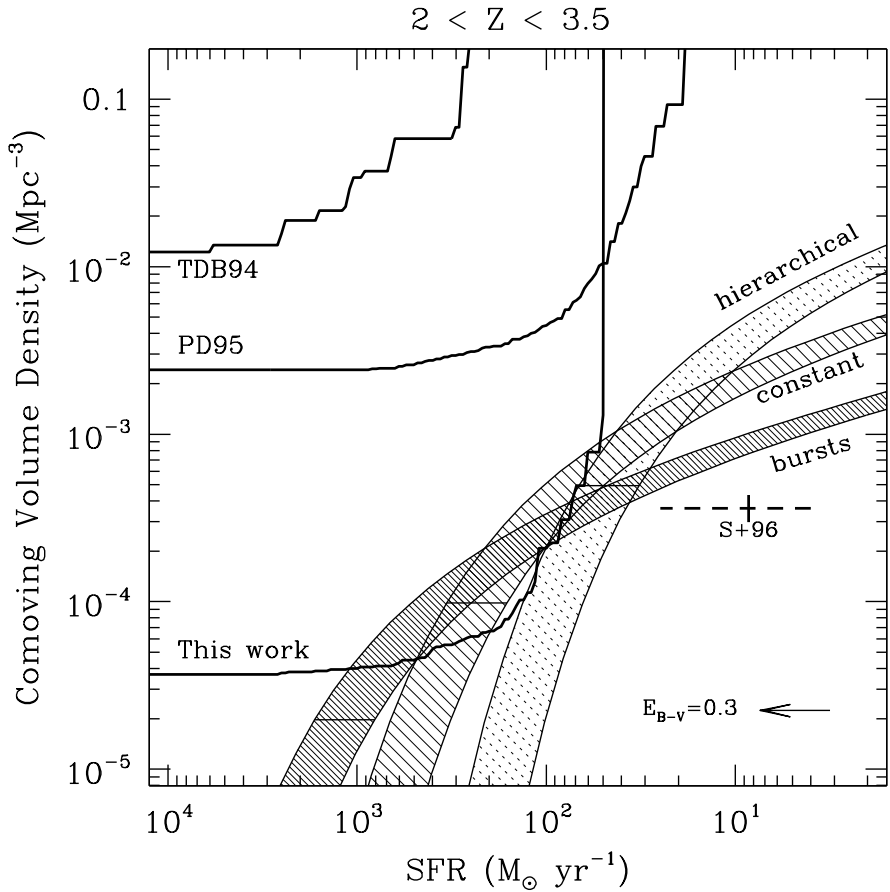}
\figcaption[Thompson.fig4.ps]{Restframe limits in the redshift range $2.0 
< z < 3.5$.  In our data, this redshift range is sampled primarily by 
H$\alpha$ and [OIII] in the $K$ band.  The three classes of models, 
descibed in the text, are labeled as {\em constant, burst} and {\em
hierarchical}.  For each class of model, we used $z_{fin} = 2$ and $3.5 
< z_{in} < 20$.  At any given SFR, the model lines indicate the expected
comoving density of objects with at least this level of star formation.  
The horizontal lines across each model region show the position of the 
L$^*$ galaxies.  The dashed line marks the comoving density 
(3.6$\times10^{-4}$ Mpc$^{-3}$) and range of SFR (4-25 M$_\odot$ yr$^{-1}$)
for the population of star-forming galaxies detected by Steidel et al. (1996) 
at $3.0 < z < 3.5$, while the mark at 8.5 M$_\odot$ yr$^{-1}$ indicates
a ``typical'' SFR as reported by the authors.  The three thick lines are 
the upper limits to the PG volume density from three surveys (this work; 
TDB94: Thompson, Djorgovski and Beckwith, 1994; PD95: Pahre and Djorgovski,
1995), where the regions to the upper right of these curves are excluded 
by the surveys.  If sufficient dust were present in the galaxies to cause 
$\sim 1^m$ extinction at visual wavelengths, the limit lines should be 
moved to the left by the amount indicated by the arrow. \label{fig4}}
\end{figure}
\clearpage

\begin{figure}
\plotone{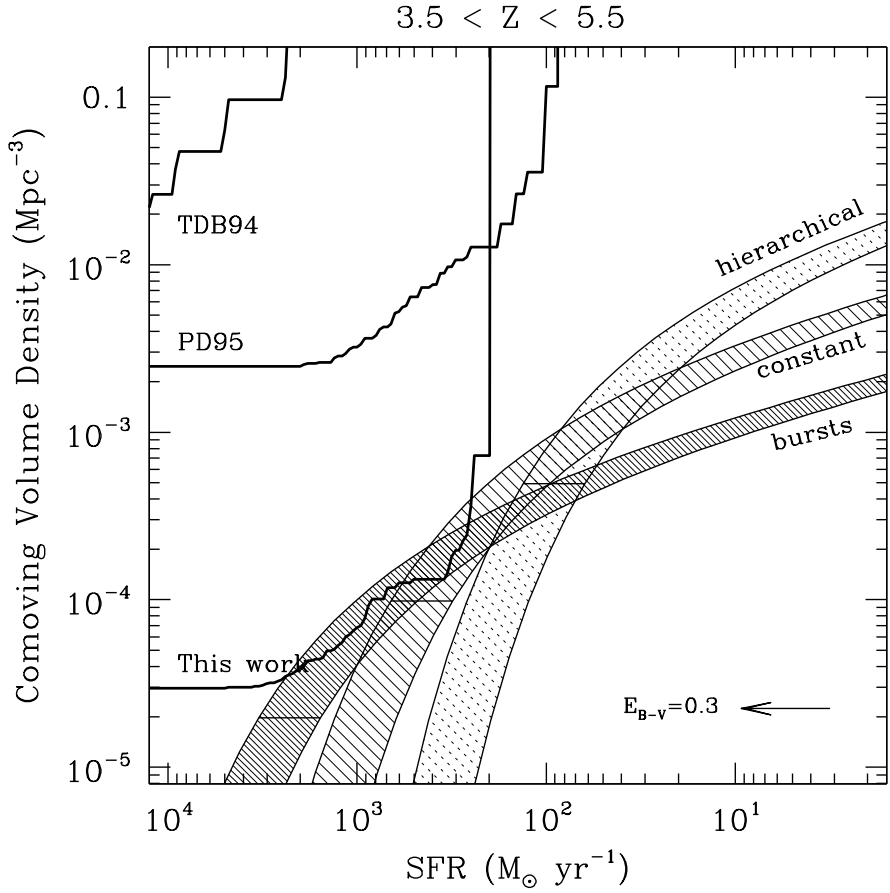}
\figcaption[Thompson.fig5.ps]{Restframe limits in the redshift range $3.5 
< z < 5.5$.  This redshift range is sampled by [O II] in the $H$ and $K$ 
bands, and by H$\beta$ and [O III] in the $K$ band.  The models are as in 
Figure 4, with $z_{fin}=3.5$ and $5.5 < z_{in} < 20$.  The effects of
dust on these bluer emission lines is again indicated by the arrow. \label{fig5}}
\end{figure}
\clearpage

\begin{figure}
\plotone{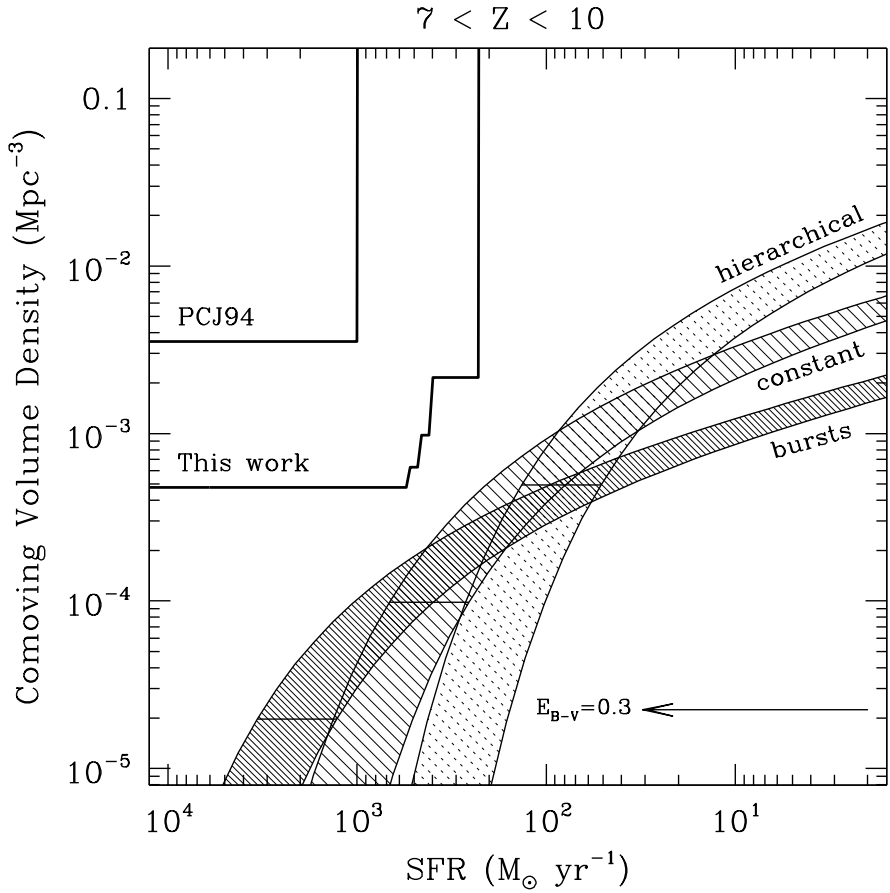}
\figcaption[Thompson.fig6.ps]{Restframe limits for Ly$\alpha$ in the $J$ 
band ($7.0 < z < 10$).  The models are the same as in Figure 4, with 
$3 < z_{fin} < 6$ and $z_{in} = 20$. For the Ly$\alpha$ line, $z_{fin}$ 
is varied rather than $z_{in}$, and $z_{fin}$ is allowed to extend to
lower redshifts, otherwise the time available at these high redshifts 
would be unreasonably short.  Dust extinction corresponding to 
$A_V \sim 1^m$, {\em neglecting resonant scattering}, is indicated by 
the arrow. \label{fig6}}
\end{figure}
\clearpage

\begin{figure}
\plotone{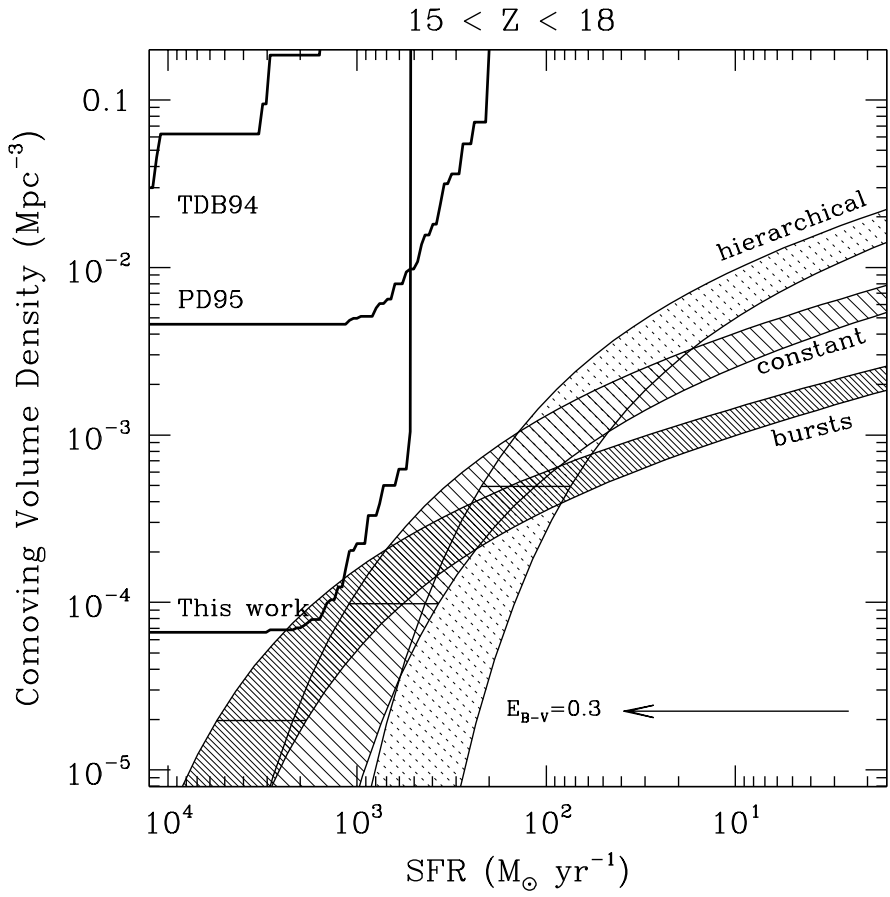}
\figcaption[Thompson.fig7.ps]{Restframe limits for Ly$\alpha$ in the $K$ 
band ($15 < z < 18$).  The models are the same as in Figure 6, with 
$4 < z_{fin} < 8$ and $z_{in} = 20$.  Extinction by dust, as in the previous figure, is indicated by the arrow. \label{fig7}}
\end{figure}
\clearpage

\end{document}